\documentclass[showpacs,aps,prl,twocolumn]{revtex4}
\usepackage{dcolumn}
\usepackage{bm}
\usepackage{graphicx}
\usepackage{amsmath}
\usepackage{latexsym}
\usepackage{amsfonts}
\usepackage{amssymb}
\usepackage{epsfig}
\usepackage{array}
\newcommand{\ket}[1]{\left\vert#1\right\rangle}
\newcommand{\ketbra}[2]{|#1\rangle \langle#2|}

\newcommand{\mm}{\,\mbox{mm}}

\newcommand{\nm}{\,\mbox{nm}}

\newcommand{\fs}{\,\mbox{fs}}

\newcommand{\MHz}{\,\mbox{MHz}}

\newcommand{\W}{\,\mbox{W}}

\begin{document}

\title{Experimental realization of Dicke states of up to six-qubits \\
for multiparty quantum networking}

\author{R. Prevedel$^1$, G. Cronenberg$^1$, M. S. Tame$^2$, M.~Paternostro$^2$, P. Walther$^{1,3}$}
\author{M. S. Kim$^2$}
\author{A. Zeilinger$^{1,3}$}
\affiliation{$^1$Faculty of Physics, University of Vienna, Boltzmanngasse 5, A-1090 Vienna, Austria\\
$^2$School of Mathematics and Physics, The Queen's University, Belfast, BT7 1NN, UK\\
$^3$Institute for Quantum Optics and Quantum Information (IQOQI),
Austrian Academy of Sciences, Boltzmanngasse 3, A-1090 Vienna,
Austria}

\date{\today}

\begin{abstract}
We report the first experimental generation and characterization of
a six-photon Dicke state. The produced state shows a fidelity of
$F\!=\!0.56\pm0.02$ with respect to an ideal Dicke state and violates a witness
detecting genuine six-qubit entanglement by four standard
deviations. We confirm characteristic Dicke properties of our
resource and demonstrate its versatility by projecting out four- and
five-photon Dicke states, as well as four-photon GHZ and W states.
We also show that Dicke states have interesting applications in
multiparty quantum networking protocols such as open-destination
teleportation, telecloning and quantum secret sharing.
\end{abstract}


\maketitle

Multipartite entanglement is at the core of studies probing the
foundations of quantum physics and represents a key component in a
wide range of quantum information processing (QIP)
tasks~\cite{Horod}.
So far, Greenberger-Horne-Zeilinger (GHZ)~\cite{ghz},
W~\cite{Wstates}, cluster and graph states~\cite{clusgraph} have
been studied and experimentally investigated~\cite{Ei}. However,
many other nonequivalent classes of quantum states with interesting
and unusual symmetries exist~\cite{class}. In particular, Dicke
states~\cite{Dicke} provide a rich opportunity for exploring
multipartite entanglement. Recent studies have focused on techniques
for generating, detecting and characterizing these
states~\cite{tothdicke,SS} in atomic, ion-trap~\cite{Agarw} and
optical~\cite{weinfurterDicke} settings.

In this Letter we report the experimental generation and
investigation of a variety of multi-photon entangled states. We
present a flexible linear-optics setup that can produce four-,
five- and six-photon representatives of the important class of Dicke
states, as well as four-photon GHZ states. Information is encoded in
the polarization degrees of freedom of entangled photons produced by
high-order spontaneous parametric down conversion (SPDC). We show
that our generated states are genuinely multipartite entangled by
using tailor-made and experimentally favorable witness tools. These
new characterization methods are important in virtue of the
non-ideal nature of the six-photon state: although spurious
nonlinear processes affect its quality, quantum features can still
be observed and characterized. We also highlight the
potential for quantum control in large Hilbert spaces by evaluating
protocols such as telecloning, open-destination teleportation and quantum
secret sharing~\cite{telemurao,Gisin,weinfurterDicke,tele,qss}.

\begin{figure}[t]
\psfig{figure=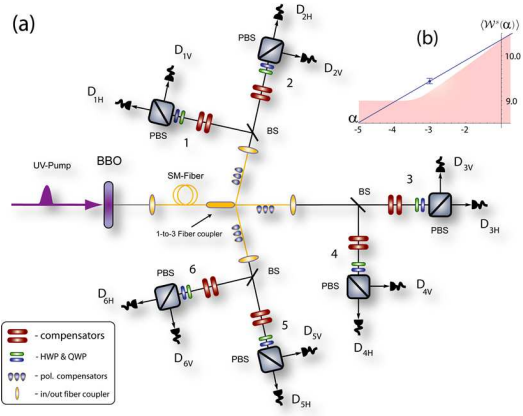,width=6.9cm} \caption{\textbf{(a)}: Setup
for the generation of the six-photon Dicke state
$|D_{6}^{(3)}\rangle$. Photons are probabilistically distributed
into modes $1,..,6$ via a 1-to-3 fiber coupler, followed by
$50\!:\!50$ BSs. The expected probability to find one photon in each
spatial mode, corresponding to the state $|D^{(3)}_6\rangle$, is
$p\sim 0.015$.
The fiber and BSs introduce birefringence, compensated by
fiber-squeezers and birefringent crystals. State-characterization is
performed via polarization-analysis of six-fold coincidences by
cascading a quarter-wave plate (QWP), half-wave plate (HWP) and a
polarizing beamsplitter (PBS), whose output ports are monitored by
multi-mode fiber-coupled single-photon detectors. Each detector
signal enters a coincidence logic that records multi-photon
coincidences. \textbf{(b)}: Biseparability region (shaded) for
$\langle {\cal W}^{s}(\alpha)\rangle_{bs}$ and experimental point
(predicted line) for $\langle {\cal
W}^{s}(\alpha)\rangle_{\varrho^{(3)}_6}$.
}
\label{fig1}
\end{figure}

{\it Experiment}.- Fig.~\ref{fig1}~{\bf (a)} shows the setup for the
generation of the three-excitation six-photon Dicke state
$|D^{(3)}_6\rangle\!=\!\frac{1}{\sqrt{20}}\sum_{P}\ket{HHHVVV}_{123456}.$ Here,~$\ket{H/V}_i$ are horizontal/vertical polarization states of a
photon in spatial mode $i=1,..,6$, which encode the logical states
of a qubit, while $\sum_{P}$ denotes the sum over all permutations
of logical states~\cite{Dicke_gen}. In the setup, six photons are
probabilistically distributed among the spatial modes by
non-polarizing beamsplitters (BSs):~upon detecting one photon in
each mode we post-selectively observe $|{D_6^{(3)}}\rangle$.
We use higher-order emissions of a collinear type-II SPDC process
for the simultaneous production of three pairs of
photons~\cite{epaps}.
A Coherent~Inc.~Verdi~V-18 laser is combined with a mode-locked Mira
HP Ti:Sa oscillator to reach the energy necessary to observe {\it
third-order} SPDC emissions.
The pulsed-laser output ($\tau=200\fs$, $\lambda=810\nm$, $76$\MHz)
is frequency-doubled using a 2\mm-thick Lithium triborate (LBO)
crystal, resulting in UV pulses of $1.4\W$~cw-average. To avoid
optical damage to the anti-reflection coating of the LBO, we
continuously translate it with a step-motor, achieving a very stable
source of UV pulses (power and count-rate fluctuations less than
$1\!-\!2\%$ over 30~h). The UV pulses are focused onto a
$2\mm$-thick $\beta$-barium borate (BBO) type-II crystal, cut for
collinear down-conversion emission. Dichroic mirrors then separate
the down-converted photons from the UV pump and a compensator erases
walk-off effects. We use high-transmittivity interference filters
($\Delta\lambda=3\nm$) to spatially and spectrally select the
down-conversion, which is coupled to a single-mode fiber guiding the
photons to the {\it Dicke-setup} of Fig.~\ref{fig1}~{\bf (a)}. At
$1.4\W$ of UV pump power, we observe~$\sim\!0.003$ six-photon Dicke
states per second. Higher power would increase the six-fold rate
while decreasing the fidelity due to undesired detection events from
higher-order SPDC emissions~\cite{epaps}.

\begin{figure}[b]
\psfig{figure=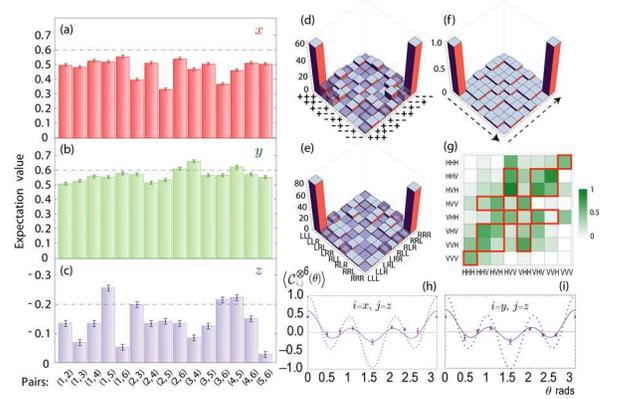,width=7.8cm}\caption{Experimental study
of the six-photon Dicke state $\rho_{6}^{(3)}$.
\textbf{(a)}-\textbf{(c)}: Correlations
$\langle \sigma_i^{(j)}\!\sigma_i^{(k)}\rangle$ for qubit pairs
$(j,k)$ ($i\!=\!x,y,z$) for $\langle{\cal W}^s(\alpha)\rangle$. Dashed
lines are ideal values. \textbf{(d)},~\textbf{(e)} and
\textbf{(g)}: Coincidences for photons measured in
$\ket{\pm}\!=\!(\ket{H}\pm\ket{V})/\sqrt{2}$,
$\ket{L/R}\!=\!(\ket{H}\pm i \ket{V})/\sqrt{2}$ and $\ket{H/V}$
(rescaled).\,\textbf{(f)}: Ideal populations for
\textbf{(d)}~and~\textbf{(e)}. \textbf{(h)}~and~\textbf{(i)}: Multiphoton correlations. Dashed (solid) lines are the patterns of $\langle{\cal
C}^{\otimes6}_{x/y,z}(\theta)\rangle$ for $|D_{6}^{(3)}\rangle$
($\rho_{\rm sim}$). The dots are experimental points.} \label{fig1b}
\end{figure}
{\it State characterization}.- In order to detect the presence of
genuine multipartite entanglement (GME) in our experimental states,
{\it i.e.} \!quantum entanglement shared by all the particles
involved, we use collective spin inequalities~\cite{SS}. Various
entanglement witnesses have been found to be well-suited to the
class of Dicke states~\cite{tothdicke}. They are experimentally
appealing, due to the small number of local measurement settings
(LMSs) required, in stark contrast to their more demanding
projector-based counterparts. We start with the collective spin
witness $\langle {\cal W}^s\rangle_f\!=\!\langle J^2_x+
J^2_y\rangle_f$, where $f$ refers to the state over which the
expectation value is calculated. Here,
$J_{i}\!=\!\frac{1}{2}\sum^N_{k=1}\sigma^{(k)}_{i}$ $(i\!=\!x,y,z)$
are collective-spin operators of $N$ qubits with label $k$ and
$\sigma^{(k)}_i$ denotes the $i$-Pauli operator. By using the
techniques described in Ref.~\cite{noteSS}, it can be seen that for
any six-qubit biseparable ($bs$) state $\langle {\cal
W}^s\rangle_{bs}\!\leq\!{11.02}$, so that $\langle {\cal
W}^s\rangle_{f}\!>\!{11.02}$ will detect the presence of GME in $f$.
{However}, due to the non-ideal two-qubit correlations upon which
${\cal W}^s$ depends (shown in Figs.~\ref{fig1b}~{\bf (a)}~and~{\bf
(b)} for $\langle J_x^2 \rangle$ and $\langle J_y^2\rangle$), our
experimental state $\varrho^{(3)}_6$ gives $\langle {\cal
W}^s\rangle_{\varrho^{(3)}_6}\!<\!11.02$. To obtain a witness {that}
detects GME for a non-ideal state, we insert a term proportional to
${J}^2_z$, for which $\langle J^2_z\rangle_{D^{(3)}_6}\!=\!0$. This
gives the more general witness ${\cal W}^s(\alpha)
\!=\!{J}^2_x+{J}^2_y+\alpha{J}^2_z$ ($\alpha\in\mathbb{R}$). We then
search for values of $\alpha$ such that $\langle{\cal
W}^s(\alpha)\rangle_{\varrho^{(3)}_6}\!>\!\langle{\cal
W}^s(\alpha)\rangle_{bs}$. In Fig.~\ref{fig1b}~{\bf (c)} we show the
two-qubit correlations for $\langle J_z^2
\rangle_{\varrho^{(3)}_6}$, which contribute to $\langle{\cal
W}^s(\alpha)\rangle$ shown in Fig.~\ref{fig1}~{\bf (b)}. A range of
$\alpha$ exists where $\langle{\cal
W}^s(\alpha)\rangle_{\varrho^{(3)}_6}\!>\!\langle{\cal
W}^s(\alpha)\rangle_{bs}$: the gap is optimized at $\alpha\!=\!-3$,
where $\langle{\cal W}^s(\alpha)\rangle_{bs}^{\rm max}-\langle{\cal
W}^s(\alpha)\rangle_{\varrho^{(3)}_6}\!=\!-0.24\pm0.06$, thus
confirming GME for our experimental state.

We now further probe the features of $\varrho^{(3)}_6$ and consider
the multi-photon correlator ${\cal C}^{\otimes
N}_{i,j}(\theta)\!=\!(\cos\theta\,\sigma_i\!+\!\sin
\theta\,\sigma_j)^{\otimes{N}}$. This allows the sampling of
$N$-photon correlations in orthogonal planes of the single-qubit
Bloch sphere, providing important information about the off-diagonal
elements of the density matrix and thus its coherence properties.
One finds $\langle {\cal C}^{\otimes 6}_{i,z}(\theta)
\rangle_{D^{(3)}_6}\!=\![3\cos(2\theta)+5\cos(6\theta)]/8$, for
$i\!=\!x,y$. Only the coherences within the  Dicke state are
responsible for the interference between the trigonometric functions
in $\langle {\cal C}^{\otimes 6}_{i,z}(\theta)
\rangle_{D^{(3)}_6}$~\cite{epaps,signature}. In
Figs.~\ref{fig1b}~{\bf (h)}~and~{\bf (i)} we compare the ideal
coherence signature with that of $\varrho^{(3)}_6$, finding a
reduced visibility. We also compare $\varrho^{(3)}_6$ with the
behavior of the state $\rho_{\text{sim}}$ resulting from a detailed
simulation of our setup including multiple-pair emissions and
losses~\cite{epaps}. The simulated state spans a Hilbert-space sector which is
larger than the $2^6$-dimensional space of $|{D^{(3)}_6}\rangle$.
Moreover, the presence of spurious state-components in
$\rho_{\text{sim}}$ affects the ideal populations and coherences, as
shown in Ref.~\cite{epaps}. The accuracy of the simulation is
confirmed by the behavior of $\langle {\cal C}^{\otimes
6}_{i,j}(\theta) \rangle_{\rho_{\text{sim}}}$ shown in
Figs.~\ref{fig1b}~{\bf (h)} and {\bf (i)}, revealing good agreement
with our data. Our analysis of $\varrho^{(3)}_6$ is strengthened by
evaluating the state fidelity $\langle F_{D^{(3)}_6}
\rangle_{\varrho^{(3)}_6}$, where the projector
$F_{D^{(3)}_6}\!=\!|D^{(3)}_6\rangle\langle{D}^{(3)}_6|$ is
decomposed into 544 terms involving Pauli operators, requiring 21
LMSs for their evaluation~\cite{LMS1}. We find $\langle
F_{D^{(3)}_6}\rangle_{\varrho^{(3)}_6}\!=\!0.56\pm0.02$, which
agrees well with the value $0.61$ from $\rho_{\text{sim}}$. The
small discrepancy is due to slightly asymmetric fiber-coupling of
$\ket{H/V}$ due to SPDC birefringence. The setup performances are
thus limited by noise from higher-order emissions~\cite{epaps}.
Despite such clearly consistent results, the measured fidelity
prevents us from unambiguously claiming that our generated state is
Dicke-class~\cite{commentoreferee}. As full state tomography is
experimentally prohibitive, we complement the fidelity analysis with
additional characterization tools.

\begin{figure}[t]
\psfig{figure=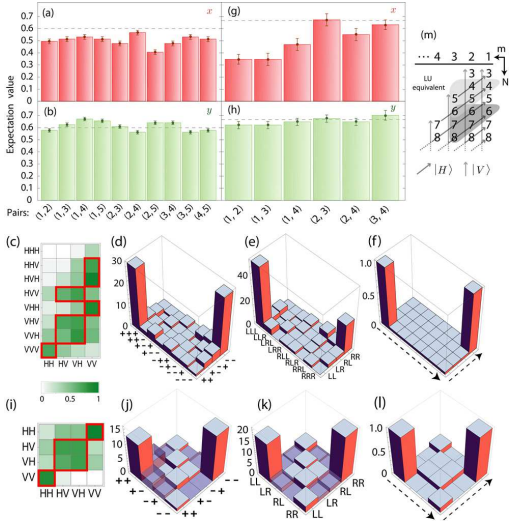,width=6.7cm} \caption{Experimental data for the Dicke
states $\varrho_{5}^{(2)}$ and $\varrho_{4}^{(2)}$.
\textbf{(a)}~and~\textbf{(b)} (\textbf{(g)}~and~\textbf{(h)}): Correlations $\langle \sigma_i^{(j)}\sigma_i^{(k)}\rangle$
for qubit pairs $(j,k)$ of $\varrho_{5}^{(2)}$
($\varrho_{4}^{(2)}$), with $i=x, y$.
\textbf{(c)}-\textbf{(e)}
(\textbf{(i)}-\textbf{(k)}): Coincidences
for the five (four) photons measured in $\ket{H/V}$, $\ket{\pm}$ and $\ket{L/R}$. \textbf{(f)}
(\textbf{(l)}): Ideal populations for
$\ket{\pm}$ and $\ket{L/R}$. \textbf{(m)}: Navigating the
Dicke class by measurement.} \label{fig2}
\end{figure}

We now explore the {\it nested} structure of Dicke states and their
persistence of entanglement by conditionally generating four and
five-photon entangled states via projections of
$|{D^{(3)}_6}\rangle$~\cite{weinfurterDicke,engineering}. For example,
by measuring one photon in $\ket{H}$, the five-photon state
$|{D^{(2)}_5}\rangle$~\cite{Dicke_gen} is projected out. This state
is equivalent to ${\sigma}^{\otimes5}_x|{D^{(3)}_5}\rangle$, showing
that {\it navigation} through the Dicke class of states is possible
via projections and local operations. Indeed, one can write
$|D^{(m)}_{N}\rangle=(C_N^m)^{-1/2}[(C_{N-1}^{m-1})^{1/2}
\ket{H}|D^{(m-1)}_{N-1}\rangle+(C_{N-1}^{m})^{1/2}
 \ket{V}| D^{(m)}_{N-1}\rangle]$ and navigate as shown in Fig.~\ref{fig2}~{\bf (m)}.
We start by experimentally projecting out the five-photon state
${\varrho^{(2)}_5}$ in modes $2,..,6$ (Figs.~\ref{fig2}~{\bf
(a)}-{\bf (f)} show the relevant experimental data).
For five-qubit states we have
$\langle {\cal W}^s\rangle_{bs}\!\leq\!{7.87}$~\cite{noteSS}, giving $\langle
{\cal W}^s\rangle_{bs}^{\rm max}-\langle {\cal
W}^s\rangle_{\varrho^{(2)}_5}\!=\!-0.21\pm0.04$, thus detecting
GME (here, $\alpha=0$ is set due to the good quality of the correlations).
To check consistency, we have also projected
photon 6 in
$\ket{H}$, finding $\langle {\cal W}^s\rangle_{bs}^{\rm max}-\langle
{\cal W}^s\rangle_{\varrho^{(2)}_5}=-0.32\pm0.02$.

\begin{figure}[t]
\psfig{figure=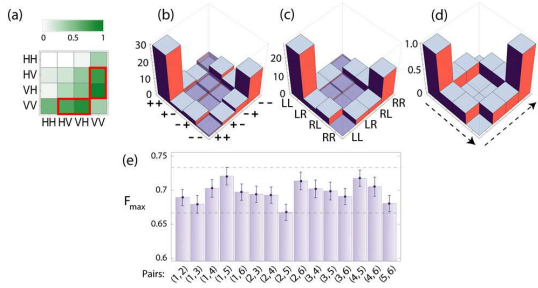,width=7.1cm} \caption{Experimental data
for the W-state
$\varrho^{(1)}_4$. \textbf{(a)}-\textbf{(c)}:
Coincidence counts for $\varrho^{(1)}_4$ in the
rescaled $\ket{H/V}$, $\ket{\pm}$ and $\ket{L/R}$ bases.
\textbf{(d)}: Ideal populations for
\textbf{(b)}~and~\textbf{(c)}. \textbf{(e)}: Maximum
achievable fidelity $F_{\text{max}}$ using pair $(i,j)$ as a
channel. The upper (lower) line shows the ideal (classical) value. }
\label{fig3}
\end{figure}

Next, we project out the four-photon Dicke state
$|D^{(2)}_4\rangle$~\cite{Dicke_gen} by measuring
one photon in $\ket{H}$ and another in $\ket{V}$. Using $\langle{\cal
W}^{s}\rangle_{bs}\leq 5.23$~\cite{tothdicke}, the correlations for
the experimental state $\varrho^{(2)}_4$ in modes $3,..,6$ (shown in
Figs.~\ref{fig2}~{\bf (g)}-{\bf (l)}) give $\langle{\cal
W}^{s}\rangle_{bs}^{\rm max}-\langle{\cal
W}^{s}\rangle_{\varrho^{(2)}_4}\!=\!-0.16\pm0.07$, thus detecting
GME. Moreover, we have evaluated the state fidelity $\langle
F_{D^{(2)}_4} \rangle_{\varrho^{(2)}_4}\!=\!0.66\pm{0.05}$ by
implementing the 9-LMS projector $F_{D^{(2)}_4}$~\cite{LMS2}. We
complete our study of four-photon Dicke states by assessing
a four-photon W state $|D^{(1)}_4\rangle$ (equivalent to
${\sigma}^{\otimes4}_{x}|D^{(3)}_4\rangle$), generated from
$|D^{(3)}_6\rangle$ upon measurement of two photons in $\ket{H}$.
Experimentally, we project the state
$\varrho^{(1)}_4$ into modes $3,..,6$ (coincidence counts
shown in Figs.~\ref{fig3}~{\bf (a)}-{\bf (d)}). Although
$|{D^{(1)}_4}\rangle$ does not exceed $\langle{\cal W}^{s}\rangle_{bs}^{\rm max}$, for our
experimental state we find $\langle{\cal W}^{s}\rangle_{bs}^{\rm
max}\!-\!\langle{\cal W}^s\rangle_{\varrho^{(1)}_4}\!=\!-0.2\pm0.1$
due to $\langle{J_{x,y}^2}\rangle$ being slightly larger than their
ideal values, thus detecting GME. We further characterize $\varrho^{(1)}_4$ by
evaluating $\langle F_{D^{(1)}_4}
\rangle_{\varrho^{(1)}_4}\!=\!0.62\pm{0.02}$, which requires only 7
LMSs~\cite{guhnepan}. Finally, a state locally equivalent to $\ket{GHZ_4}\!=\!(1/\sqrt
2)[\ket{H}^{\otimes4}+\ket{V}^{\otimes4}]$ can also be generated
from $|{D^{(3)}_6}\rangle$ by measuring one photon in $\ket{+}$ and
another in $\ket{-}$. Ideally, this produces
$(|{D^{(1)}_4}\rangle\!-\!|{D^{(3)}_4}\rangle)/\sqrt{2}\equiv{\sigma}^{(1)}_z({\cal
H}\sqrt{{\sigma_z}})^{\otimes4}\ket{GHZ_4}$ ($\cal{H}$ is the
Hadamard gate). The state fidelity (using 5 LMSs) is
$\langle F_{GHZ_4} \rangle_{\varrho_{GHZ}}\!=\!0.56\pm{0.02}$, giving
a projector-based witness value of $\langle {\cal
W}\rangle_{GHZ} =-0.06\pm0.02$~\cite{guhnepan}, thus confirming again GME.

{\it Quantum protocols}.- Despite the non-ideal value of the
state-fidelity, the symmetries within our six-photon resource make it suitable for several key quantum
networking protocols~\cite{weinfurterDicke,tele}, some of which have been demonstrated in four-photon settings. Tracing out $N\!-\!2$ qubits, one finds the
two-photon state $\rho=\alpha_N
\ketbra{\psi^+}{\psi^+}+(1-\alpha_N)[\ketbra{HH}{HH}+\ketbra{VV}{VV}]/2$
with $\alpha_N=N/[2(N-1)]$ for $N\ge4$~\cite{SS} and
$\ket{\psi^{\pm}}=(\ket{HV}\pm\ket{VH})/\sqrt{2}$. Here, the maximal
singlet fraction $F_{\rm msf}$~\cite{Horo}, given by the maximum of
$\langle F_{\psi^-}\rangle$ under local operations and classical
communication, helps in assessing the usefulness of $\rho$ for
networking tasks. We consider using $\rho$ as a teleportation
channel~\cite{tele,weinfurterDicke}, where the maximum fidelity
achievable for teleporting an arbitrary state is
$F_{\text{max}}\!=\!(2F_{\text{msf}}+1)/3$. For
$|D_{N}^{(N/2)}\rangle$, $F_{\rm msf}=\alpha_N$, thus
$F_{\text{max}}\!=\!\frac{2N-1}{3(N-1)}$. Fig.~\ref{fig3}~{\bf (e)}
shows $F_{\text{max}}$ for all pairs of photons from
$\varrho^{(3)}_6$ and the ideal value $0.73$ (upper dashed-line). As
any photon pair in $|D_{6}^{(3)}\rangle$ provides a channel for teleportation, regardless of operations applied to the others,
one can use it for
telecloning~\cite{telemurao,Gisin,weinfurterDicke}. The
fidelity-limit for universal symmetric $1\!\to\!(N\!-\!1)$ cloning
is exactly $F_{\text{max}}$~\cite{Gisin}, thus
$|D_{N}^{(N/2)}\rangle$ is an ideal resource for this task.
Following~\cite{telemurao}, we have evaluated the protocol using
$\varrho^{(3)}_6$: Fig.~\ref{fig3}\,{\bf (e)} shows that the maximum
cloning fidelity achievable is consistently above the classical
threshold of $2/3$~\cite{Gisin}. A perfect $\ket{\psi^+}$ channel
for teleportation can be created, with success probability
$\alpha_N$, if $N\!-\!2$ photons are {\it measured} out of
$|D_{N}^{(N/2)}\rangle$ in the $\ket{H/V}$ basis. This is in
contrast to telecloning, where the photons are traced-out, resulting
in an imperfect channel with fidelity $\langle F_{\psi^+}
\rangle_{\rho}=\alpha_N$. As the core operation needed for
telecloning commutes with the $\ket{H/V}$ measurements, one can
choose between telecloning and teleportation~\cite{weinfurterDicke},
with the success probability to teleport to any one party given by
$p_s\!=\!\alpha_N\!\ge\!\frac{1}{2}$. Thus $|D_{6}^{(3)}\rangle$ can
be used for open destination
teleportation~\cite{tele,weinfurterDicke}. For $\varrho^{(3)}_6$ we
find a mean value $\bar{p}_s=0.55\pm0.02$ very close to the ideal
${p}_s\!=\!0.6$. As an example, we choose photons 5 and 6, finding a
mean fidelity $\langle \bar{F}_{\psi^+} \rangle_{\rho_{exp}}=0.71
\pm 0.02$.

Finally, $|D_{N}^{(N/2)}\rangle$ can also be used for multiparty
quantum secret sharing~\cite{qss}, where entanglement ensures that
all parties must cooperate in order to obtain a shared secret. The
trick is to exploit the perfect correlations in the
maximally-conjugate bases of ${\sigma}_{x,y}$. Using $\langle {\cal
C}^{\otimes N}_{i,j}(\theta) \rangle$, we get $\langle
\sigma_{x,y}^{\otimes N} \rangle_{D_{N}^{({N}/{2})}}\!=\!1$.
Consider the $\sigma_{x}$ basis and $x_j\!\in\!\{0,1\}$ as the
measurement-outcome for the $j$-th photon. If photon $1$ is
measured, the value of $x_1$ can only be recovered via
$x_1=\oplus_{i=2}^{N} x_i$ ($\oplus$ denotes mod-2 addition),
implying cooperation of parties $2,..,N$. As $|D_{N}^{(N/2)}\rangle$
is symmetric, this applies to any choice for the initial party. The
same holds for the ${\sigma}_y$ basis. When the parties announce
their (randomly chosen) bases, a shared-key can be
distributed~\cite{qss}, with which a designated party encodes the
secret. Any set of less than $N\!-\!1$ parties cannot recover the
key and although subsets of parties can exist with partial
information about $x_1$ (for instance), any such bias is removable by
post-processing~\cite{qss,Weinqss}. We thus evaluated the
{expected} quantum bit error-rate of the generated key (before
post-processing), given by the average error rate of the
${\sigma}_{x,y}$ bases. We find $25\pm 2 \%$ and $29\pm 1 \%$ for
$N=4$ ($415$ shared bits) and $N=6$ ($889$ shared bits)
respectively over an $82$ hour period.

{\it Remarks}.- We have demonstrated a linear-optics setup able to
produce various states from the Dicke class and characterized
their properties using new methods. We also
evaluated the potential of our six-photon state for multiparty
quantum networking. Our work significantly extends the range of
attainable quantum states and paves the way toward the experimental
study of other six-qubit entangled states~\cite{class} (and larger ones) and
their use in QIP.

{\it Acknowledgments}.- We acknowledge discussions with J.~Kofler, T.~Jennewein
and N.~Langford, help from B.~Rauer and support from EPSRC, QIPIRC,
FWF, EC under the Integrated Project Qubit Application, EMALI and
the U.S.\,Army\,Research\,Funded IARPA.


\begin{thebibliography}{99}
\bibitem{Dicke_gen} A Dicke state of $N$ qubits and  $k$ excitations is given by $|D_N^{(k)}\rangle\!=\!(C_N^k)^{-1/2}\sum_{P} \ket{H}^{\otimes k}\ket{V}^{\otimes
N-k}$, with $C_i^j$ as the binomial coefficient.
\bibitem{commentoreferee} In the $\sigma_{x,y}$ bases, for instance, $|D^{(3)}_6\rangle$ has an overlap of $\sqrt{5/8}$ with a six-photon GHZ state.
\end{thebibliography}
\end{document}